\documentclass[12pt, prd, showpacs]{revtex4}
\usepackage{amsmath}
\usepackage{amssymb}

\input{tcilatex}

\begin{document}

\title{Truly naked spherically-symmetric and distorted black holes}
\author{O. B. Zaslavskii}
\affiliation{Astronomical Institute of Kharkov V. N. Karazin National University,
Ukraine, Svoboda Square 4, Kharkov 61077, Ukraine}
\email{ozaslav@kharkov.ua}

\begin{abstract}
We demonstrate the existence of spherically-symmetric truly naked black
holes (TNBH) for which the Kretschmann scalar is finite on the horizon but
some curvature components including those responsible for tidal forces as
well as the energy density $\bar{\rho}$\ measured by a free-falling observer
are infinite. We choose a rather generic power-like asymptotics for the
metric functions and analyze possible types of a horizon depending on the
behavior of curvature components in the free-falling frame. It is also shown
in a general case of distorted black holes that $\bar{\rho}$ and tidal
forces are either both finite or both infinite. The general approach
developed in the article includes previously found examples and, in
particular, TNBHs with an infinite area of a horizon. The fact that the
detection of singularity depends on a frame may be relevant for a more
accurate definition of the cosmic censorship conjecture. TNBHs may be
considered as a new example of so-called non-scalar singularities for which
the scalar curvature invariants are finite but some components of the
Riemann tensor may diverge in certain frames.
\end{abstract}

\maketitle



\section{Introduction}

Usually, regular or singular character of points or surfaces in spacetime
reveals itself as an inner property inherent to the manifold and does not
depend on the frame in which it is described. In particular, the value
Kretschmann scalar is finite or infinite, whatever frame be used for its
calculation. Nonetheless, as was pointed out in \cite{nk1}, \cite{nk2}, in
the vicinity of black holes regular and singular features may entangle in a
non-trivial way. It turns out that in some cases the curvature components in
the free-falling frame are enhanced significantly with respect to their
static values to the extent that they are finite non-zero in spite of in the
static frame they are negligible. The similar observation was made for
particular types of black holes with an infinite horizon area in some
Branse-Dicke theories \cite{br1}, \cite{br2}. The reason lies in the
singular character of the static frame itself: on the horizon a time-like
static observer becomes null and the frame by itself fails. The term "naked
black holes" was suggested in \cite{nk1}, \cite{nk2} for such objects.
Strictly speaking, the word "naked" is not quite exact here since all
curvature components remain finite in the free-falling frame. One may ask
whether it is possible to make the next step and find the horizons for which
the distinction between both frames is even more radical in the following
sense. In the static frame all curvature components are finite but in the
free-falling one some of them as well as corresponding tidal forces diverge.
The answer is positive \cite{v}. The explanation how to reconcile some
infinite components of the Riemann tensor in the orthonormal free-falling
frame with the finiteness of the Kretschmann scalar is connected with the
Lorentz signature of the spacetime. In the static frame all components of
the curvature enter the Kretschmann scalar with the same sign but in the
free-falling one this is not the case. As a result, different divergent
terms cancel each other and the net outcome is finite. (From a more general
viewpoint, the systems discussed in our paper represent examples of
so-called non-scalar polynomial curvature singularities \cite{he} (Chap. 8), 
\cite{es}, as it will be clear below.) To distinguish black holes with
infinite tidal forces on the horizon in the free-falling frame from "naked"
ones \cite{nk1}, \cite{nk2} the term "truly naked black holes" (TNBH) was
suggested in \cite{v}. It is also worth noting that the crucial difference
between a static frame and a free-falling one on the horizon reveals itself
not only in the dynamic effects (such as tidal force) but also in the
algebraic structure of the gravitational field. Thus, the Petrov type of the
field on the horizon detected by a free-falling observer can differ from the
limit measured by a sequence of static observers in the near-horizon limit 
\cite{v}.

TNBH considered in \cite{v} are entirely due to distorted, non-spherical
character of the metric. In doing so, certain Weyl scalar (quantities
obtained by the projection of the Weyl tensor to the null tetrad attached to
an observer) diverge on the horizon \cite{v}. Meanwhile, the full set of
quantities that determines gravitational field includes, apart from Weyl
scalars, also components of the Ricci tensor. In cases of distorted TNBH
considered in \cite{v} both types of quantities diverge on the horizon. In
general, this is not necessarily so. In the spherically-symmetrical case
there is only one non-vanishing Weyls scalar (this is is connected with the
fact that a generic spherically-symmetric gravitational field is of type D 
\cite{w}). Its value is finite and coincides in both frames (static and
free-falling). Then, the only potential source of divergencies is the Ricci
tensor in a free-falling frame and the corresponding components of the
stress-energy tensor. Therefore, there remains open window for the existence
of TNBH even in the spherically-symmetrical case in spite of the finiteness
of Weyl scalars.

The aim of the present paper is to fill this gap and show that
spherically-symmetrical TNBH do exist. We demonstrate that this is possible
in the case of extremal and so-called ultra-extremal horizons whereas for
the non-extremal case we return to the situation already discussed in \cite%
{nk1}, \cite{nk2}. As the key role is played by the asymptotic behavior of
the metric near the horizon, we suggest a general approach which relies on
this asymptotics only and does not require the knowledge of the metric
everywhere. The corresponding general approach includes cases considered in 
\cite{nk1} - \cite{br2}. We also enlarge previous analysis carried out in 
\cite{v}. We analyze explicitly the behavior of all curvature components for
non-extremal, extremal and ultra-extremal horizons and conclude that
divergencies in tidal transverse forces is the sufficient criterion to
include an object to the class of TNBHs. Apart from this, we also show for
the generic distorted horizon that divergencies of the energy density in the
free-falling frame may be compatible with the finiteness of the Kretschmann
scalar.

\section{Spherically-symmetrical case}

\subsection{Static frame}

Let us consider the spherically-symmetric metric%
\begin{equation}
ds^{2}=-dt^{2}U+V^{-1}dr^{2}+r^{2}(d\theta ^{2}+\sin ^{2}\theta d\varphi
^{2})
\end{equation}%
supported by the stress-energy tensor having the form%
\begin{equation}
T_{\mu }^{\nu }=diag(-\rho \text{, }p_{r}\text{, }p_{\perp }\text{, }%
p_{\perp })\text{.}
\end{equation}

It follows from $00$ and $rr$ Einstein equations that%
\begin{equation}
U=V\exp (2\psi )\text{,}  \label{uv}
\end{equation}%
\begin{equation}
\psi =4\pi \int^{r}d\bar{r}\bar{r}Q\text{, }Q=\frac{(p_{r}+\rho )}{V}\text{.}
\label{psi}
\end{equation}%
The non-vanishing components of the curvature tensor read%
\begin{equation}
R_{0r}^{0r}=E=-\frac{V^{\prime }\Phi ^{\prime }}{2}-V(\Phi ^{\prime \prime
}+\Phi ^{\prime 2})\text{, }U\equiv \exp (2\Phi )\text{,}
\end{equation}%
\begin{equation}
R_{0\theta }^{0\theta }=\bar{E}=-\frac{V\Phi ^{\prime }}{r}\text{, }R_{\phi
\theta }^{\phi \theta }=F=\frac{1-V}{r^{2}}
\end{equation}%
\begin{equation}
R_{r\theta }^{r\theta }=\bar{F}=-\frac{V^{\prime }}{2r}\text{.}
\end{equation}%
For what follows, it is convenient to introduce the quantity%
\begin{equation}
Z=\bar{F}-\bar{E}=\frac{\psi ^{\prime }V}{r}=4\pi QV=4\pi (p_{r}+\rho )\text{%
.}  \label{zr}
\end{equation}

\subsection{Free-falling frame}

Consider also the boosted frame moving in the radial direction with the
boost angle $\alpha $. Then the curvature components (hat stands for the
orthonormal frame) are transformed according to \cite{nk1} 
\begin{equation}
\hat{R}_{0^{\prime }r^{\prime }0^{\prime }r^{\prime }}=\hat{R}_{0r0r}\text{, 
}\hat{R}_{0^{\prime }\theta ^{\prime }r^{\prime }\theta ^{\prime }}=-\cosh
\alpha \sinh \alpha Z\text{, }  \label{tr1}
\end{equation}%
\begin{equation}
\hat{R}_{0^{\prime }\theta ^{\prime }0^{\prime }\theta ^{\prime }}=\hat{R}%
_{0\theta 0\theta }+\sinh ^{2}\alpha Z,\hat{R}_{r^{\prime }\theta ^{\prime
}r^{\prime }\theta ^{\prime }}=\hat{R}_{r\theta r\theta }+\sinh ^{2}\alpha Z
\label{tr2}
\end{equation}%
and similarly for components with $\theta $ replaced by $\phi $. Here $\cosh
\alpha =\frac{\varepsilon }{\sqrt{N}}$, $\varepsilon $ is the energy per
unit mass. (We choose $\alpha >0$, then our definition differs from that in 
\cite{nk1} by the sign.)

As the difference between the static and boosted frame reveals itself for
all components (except $0r0r$ one ) in a similar way, the analysis in \cite%
{nk1} was mainly restricted to the component $R_{0\theta 0\theta }$ that has
a clear physical meaning, being responsible for tidal forces in the
transverse directions. It is somewhat more convenient to deal with the
combination of two components $Z$ that includes the effect of tidal forces.
From geodesics equations, one can obtain easily (cf. \cite{2d} for the
two-dimensional analogue) that the quantity $Z$ is related to the energy
density of the source $\bar{\rho}$ measured by a free-falling observer:%
\begin{equation}
\bar{\rho}=T_{\mu \nu }u^{\mu }u^{\nu }=\frac{\varepsilon ^{2}Z}{4\pi U}%
-T_{r}^{r}(1+\frac{L^{2}}{r^{2}})+\frac{T_{\phi }^{\phi }L^{2}}{r^{2}}\text{.%
}  \label{er}
\end{equation}%
$\varepsilon $ is the energy of a particle per unit mass along the
geodesics. It follows from the transformation laws (\ref{tr1}), (\ref{tr2})
that the quantity $Z$ transforms as 
\begin{equation}
\tilde{Z}=Z\,(2\frac{\varepsilon ^{2}}{U}-1)=8\pi Q\varepsilon ^{2}\frac{V}{U%
}-Z  \label{z}
\end{equation}%
(see also derivation in a more general case of an arbitrary static metric in
Sec. IV below). Thus,%
\begin{equation}
\bar{\rho}=\frac{\bar{Z}+Z}{8\pi }-T_{r}^{r}(1+\frac{L^{2}}{r^{2}})+\frac{%
T_{\phi }^{\phi }L^{2}}{r^{2}}\text{.}  \label{rrz}
\end{equation}

It follows from (\ref{rrz}) that $\bar{\rho}$ diverges on the horizon if and
only if when $\bar{Z}$ does so. In other words, on the horizon $\bar{\rho}$
is infinite for TNBH and finite for usual and naked black holes.

We suppose that there is a horizon at $r=r_{0}$. We restrict ourselves by
the simple asymptotics%
\begin{equation}
V\thickapprox a(r-r_{0})^{p}\text{, }U\thickapprox b(r-r_{0})^{q}  \label{pq}
\end{equation}%
(more general discussion of the behavior of the metric functions near the
regular horizon can be found in \cite{gap}). As $r=r_{0}$ corresponds to the
horizon, we must have $q>0$, $p>0$, as usual.

It follows from the finiteness of $\bar{E}$ and $\bar{F}$ on the horizon
that $p\geq 1$. Then, $E$ is the only potentially diverging term in the
vicinity of the horizon:%
\begin{equation}
E=-\frac{aq(r-r_{0})^{p-2}}{4}(p+q-2)+O((r-r_{0})^{p-1})\text{,}  \label{e}
\end{equation}

If $p=1$, the regularity of the spacetime selects the only value $q=1$. If $%
p\geq 2$, the geometry is regular on the horizon for any $q$.

Then, by direct substitution, we obtain that near the horizon%
\begin{equation}
Z\sim (q-p)(r-r_{0})^{p-1}\text{, }\tilde{Z}\sim (q-p)(r-r_{0})^{p-1-q}\text{%
,}  \label{zdv}
\end{equation}%
\begin{equation}
Q\thickapprox \frac{(q-p)}{4\pi r_{0}(r-r_{0})}+Q_{0}  \label{qp}
\end{equation}%
where $Q_{0}$ is a constant. If $p>1$, we obtain at once from (\ref{zdv})
that $Z\rightarrow 0$ on the horizon. If $p=1=q$, the leading terms of the
order unity mutually cancel in (\ref{zdv}), so that the main contribution
has the order $r-r_{0}$ and, again, $Z\rightarrow 0$ on the horizon.

In what follows, we also need the proper time of motion (for time-like
geodesics) or affine parameter (for light-like ones). In both cases we will
denote it as $\tau $. Then, from the equations of motions and conservation
of energy we have%
\begin{equation}
\tau =\int \frac{dr}{\sqrt{Y}}\text{, }Y=V(\frac{E^{2}}{U}-\frac{L^{2}}{r^{2}%
}+\delta )
\end{equation}%
where $L=u_{\phi }$ is the angular momentum, $\delta =-1$ for time-like
geodesics and $\delta =0$ for light-like ones.

In our case the time needed to reach the horizon diverges as%
\begin{equation}
\tau \sim (r-r_{0})^{c}\text{, }c=\frac{2+q-p}{2}\text{, }p-q>2\text{, }
\label{c}
\end{equation}%
or%
\begin{equation}
\tau \sim -\ln (r-r_{0})\text{, }p-q=2
\end{equation}%
and is finite if $p-q<2$.

We want to examine under what conditions 1) the horizon is regular, 2)
elucidate its nature. As far as point 2) is concerned, we distinguish three
cases:

(i) $Z\rightarrow 0$, $\tilde{Z}\rightarrow 0$ (by definition, "usual"),
(ii) $Z\rightarrow 0$, $\tilde{Z}\rightarrow const\neq 0$ ("naked"), (iii) $%
Z\rightarrow 0$, $Z\rightarrow \infty $ ("truly naked").

According to p. 1), we want to eliminate leading and subleading divergencies
in (\ref{e}). If $p<2$, we must choose $p+q=2$. For the subleading
divergencies to be absent, we must also choose $p=1$ if, as usual, only
integer power degree are allowed (in some special cases this is not
necessary if the coefficient at the term of the order $p-1$ also vanishes
but this depends strongly on the details of a system and we do not discuss
this case), so that $q=1$. If, for generality, not-integer $p$ are also
allowed, one may take $\,1<p<2$ and $q=2-p$. Then, $\tilde{Z}\sim
(r-r_{0})^{2p-3}$. In all other cases we assume that $p\geq 2$ (cases 5 - 10
below). By definition, $p=2$ represents the extremal case and $p>2$
corresponds to the ultraextremal one.

Then, the set of possibilities can be collected in table 1.

Table 1. Types of horizons with finite area.

\begin{tabular}{|l|l|l|l|l|}
\hline
&  & Type of horizon & $Q$ & $\tau $ \\ \hline
1 & $p=q=1$ & usual ($Q=0$) or naked ($Q\neq 0$) & finite & finite \\ \hline
2 & $1<p<\frac{3}{2}$ & truly naked & infinite & finite \\ \hline
3 & $p=\frac{3}{2}$ & naked & infinite & finite \\ \hline
4 & $\frac{3}{2}<p<2$ & usual & infinite & finite \\ \hline
5 & $p<q$ & truly naked & infinite & finite \\ \hline
6 & $p=q\geq 2$ & usual ($Q=0$) or naked ($Q\neq 0$) & finite & finite \\ 
\hline
7 & $q<p<q+1$ & truly naked & infinite & finite \\ \hline
8 & $p=q+1$ & naked & infinite & finite \\ \hline
9 & $q+1<p<q+2$ & usual & infinite & finite \\ \hline
10 & $p\geq q+2$ & usual & infinite & infinite \\ \hline
\end{tabular}

\smallskip \allowbreak

The Schwarzschild and Reissner-Nordstr\"{o}m black holes belong to class 1)
with $Q=0$, whereas the examples considered in \cite{nk1}, \cite{nk2} fall
into the same class with $Q\neq 0$. Case 6) includes, for example, the
Reissner-Nordstr\"{o}m - de Sitter ultracold horizon \cite{rom}. In cases 2)
- 5), 7) - 10) the quantity $Q$ is infinite on the horizon but, nonetheless,
the horizon is regular. Moreover, in cases 4), 9) and 10) the horizon is
usual in spite of divergencies in $Q$. Even if $\psi ^{\prime }\sim Q$ is
infinite on the horizon, in the product $\psi ^{\prime }V$ the second factor
overcomes divergencies in all cases, so that $Z=0$ and $p_{r}+\rho =0$ on
any regular spherically-symmetrical horizons - not only non-extremal but
also extremal and ultraextremal ones. For comparison, it is instructive to
mention that for the distorted extremal TNBHs the analog of the latter
equality does not in general hold that violates the horizon structure of the
Enistein tensor \cite{v} \ typical of usual (not naked) black holes \cite%
{vis}. In cases 2), 5) and 7) the energy density $\bar{\rho}$ measured by a
free-falling observer is infinite but the horizon is regular (naked or truly
naked).

It is worth noting that the parameter $\tau $ needed to reach the horizon is
finite except case 10. This can occur for usual horizons only. Then, the
horizon is at infinite proper distance and $\tau $ is also infinite. In this
sense, this is null infinity rather than the horizon. If only integer $p$
and $q$ are considered (as it takes place usually), $\ $cases 2) - 4), 7)
and 9) are absent.

\subsection{Behavior of tidal forces and exotic matter}

The equality $p_{r}+\rho =0$ which has to be satisfied on the horizon means
that the null energy condition (NEC)\ is satisfied on the verge. In recent
years, there is big interest to systems for which NEC is violated ($%
p_{r}+\rho <0$) since, from one hand, such a kind of matter leads to
acceleration of Universe \cite{ac1}, \cite{ac2} and, from the other hand, is
necessary ingredient for the existence of wormholes \cite{th}, \cite{book}.
In the present context, it follows from (\ref{zr}) that the validity or
violation of NEC are determined by the same quantity $Z$ that determines
also the behavior of tidal forces and%
\begin{equation}
p_{r}+\rho \sim (q-p)(r-r_{0})^{p-1}\text{.}
\end{equation}

Thus, if $q<p$, there is a region adjacent to the horizon inside which the
matter is exotic. According to table 1, TNBH can occur both with the exotic (%
$p_{r}+\rho <0$, case 7) and normal ($p_{r}+\rho >0$, case 5) matter in the
vicinity of the horizon in the external region.

\section{Infinite area of horizon}

It was tacitly assumed in the above consideration that, as usual, the
horizon radius $r_{0}$ is finite. Meanwhile, there are cases when this
condition is violated and the asymptotics of metric functions has the form

\begin{equation}
U\sim r^{q}\text{, }V\sim r^{p}  \label{pqr}
\end{equation}%
where $q<0$ and $p$ are not necessarily integers, $r\rightarrow \infty $.
For example, it occurs for some kinds of wormholes supported by phantom
matter, with $q=-1$ and $p=0$ \cite{ph} or black holes in Brans-Dicke theory
(see below). In doing so,

\begin{equation}
Z\sim \bar{F}\sim r^{p-2}\sim \bar{E}\sim E\text{,}
\end{equation}%
\begin{equation}
F\sim r^{p-2}\text{, }p\geq 0\text{ or }r^{-2}\text{, }p<0
\end{equation}%
\begin{equation}
\bar{Z}\sim r^{-2c}\text{, }
\end{equation}%
where $c$ is defined in (\ref{c}).

The system is regular if $p\leq 2$, then the proper distance to such a
"horizon" $l$ is infinite. Repeating our analysis we obtain the following
set of possible cases.

Table 2. Types of horizons with infinite area.

\begin{tabular}{|l|l|l|l|l|l|}
\hline
&  & $Z$ & $\tilde{Z}$ & Type of horizon & $\tau $ \\ \hline
1 & $p=2$ & finite & infinite & truly naked & finite \\ \hline
2 & $2-\left\vert q\right\vert <p<2$ & $0$ & infinite & truly naked & finite
\\ \hline
3 & $p=2-\left\vert q\right\vert $ & $0$ & finite & naked & infinite \\ 
\hline
4 & $p<2-\left\vert q\right\vert $ & $0$ & $0$ & usual & infinite \\ \hline
\end{tabular}

The fact that $Z$ does not vanish on the horizon in case $p=2$ is due to an
infinite area because of which dependence of $Z$ on $r$ changes as compared
to the case of finite $r_{0}$ considered in the previous Section. In case 3
and 4 a naked horizon is combined with an infinite proper time. Then, an
observer never reaches it and the horizon reveals itself as null infinity
rather a horizon in its strict sense.

Now, for illustration, we remind here two examples of exact solutions in
Brans-Dicke theory \cite{j}, \cite{bd}, \cite{br1}, \cite{br2}, \cite{cl}.
As the theory differs from general relativity, eqs. (\ref{uv}), (\ref{psi})
are not longer valid since they rely on Einstein equations but the rest of
formulas retains their meaning.

1) 
\begin{equation}
ds^{2}=-(1-\frac{\rho _{+}}{\rho })^{Q-\chi }dt^{2}+(1-\frac{\rho _{+}}{\rho 
})^{-Q}d\rho ^{2}+\rho ^{2}(1-\frac{\rho _{+}}{\rho })^{1-Q}(d\theta
^{2}+\sin ^{2}\theta d\phi ^{2})\text{.}
\end{equation}%
Let $\rho \rightarrow \rho _{+}$, then $l\rightarrow \infty $, $r\sim \rho
_{+}(1-\frac{\rho _{+}}{\rho })^{\frac{1-Q}{2}}$. The horizon exists, if $%
Q>\chi $. If $Q>1$, the area of the horizon at $\rho =\rho _{+}$ is
infinite. Regularity demands that $Q\geq 2$. Now

$\left\vert q\right\vert =\frac{2(Q-\chi )}{Q-1}$, $p=\frac{2}{Q-1}$. Then,
according to the table, we have a truly naked black hole ($Q\geq 2$, $\chi
<2 $), naked ($Q>2$, $\chi =2$) or usual ($Q>$ $\chi >2$)$.$

2) 
\begin{equation}
ds^{2}=e^{-su}[-e^{-2bu}dt^{2}+\frac{e^{2bu}}{u^{2}}(\frac{du^{2}}{u^{2}}%
+d\omega ^{2})]\text{, }s\neq 0\text{.}
\end{equation}%
This case combines a power-like and exponential asymptotics, so the analysis
is slightly modified. Spatial infinity corresponds to $u=0$. Regular horizon
exists at $u\rightarrow +\infty $ if

\begin{equation}
b>0\text{, }-2b<s<2b\text{.}
\end{equation}%
As the quantities $b$, $s$ are related to the Brans-Dicke parameter $\omega $
according to $s^{2}(\omega +\frac{3}{2})=-2b^{2}$ (see the aforementioned
references for details), it is implied that $s\neq 0$. It was noticed in 
\cite{br1}, \cite{br2} that $\tau \sim \int \frac{du}{u^{2}}\exp (-su)$ may
be infinite depending on the parameter $s.$ We want to add to this
observation the relationship between $Z$ and $\tilde{Z}$. Then,

\begin{equation}
Z\sim \exp [u(s-2b)]\rightarrow 0\text{, }\tilde{Z}\sim \exp (2su)
\end{equation}%
and we obtain the following set of variants - see Table III.

Table 3. Types of horizons in Brans-Dicke theory (case 2).

\begin{tabular}{|l|l|l|l|l|l|}
\hline
&  & $Z$ & $\tilde{Z}$ & Type of horizon & $\tau $ \\ \hline
1 & $s<0$ & $0$ & $0$ & usual & infinite \\ \hline
2 & $s>0$ & $0$ & infinite & truly naked & finite \\ \hline
\end{tabular}

Thus, examples considered in \cite{br1}, \cite{br2} blend with our general
scheme. It was pointed out in \cite{br1}, \cite{br2} that in corresponding
black hole solutions infinite tidal forces appear only with the combination
with finite proper time. It is seen from tables 1-3 that such a combination
persists in a general case of black holes with a finite or infinite horizon
area and the behavior of metric functions described by (\ref{pq}) or (\ref%
{pqr}).

\section{Distorted horizons}

The generic static metric may be written in the Gauss normal coordinates as

\begin{equation}
ds^{2}=-N^{2}dt^{2}+dn^{2}+\gamma _{ab}dx^{a}dx^{b}\text{,}  \label{gd}
\end{equation}%
where $x^{1}=n$, $a,b=2,3$. The horizon corresponds to $N=0$.

Let an observer move in the $n$ direction only. We attach the tetrad to him
that includes the vector of the four-velocity $u^{\mu }$, the vector $e^{\mu
}$ lying in the $t-n$ submanifold and orthogonal to it. For a static
observer, $u^{\mu }=N^{-1}(1,0,0,0)$, $e^{\mu }=(0,1,0,0)$. If an observer
moves along the geodesics with the energy $\varepsilon $ per unit mass, the
local Lorentz boost reads 
\begin{equation}
\bar{u}^{\mu }=u^{\mu }\cosh \alpha -e^{\mu }\sinh \alpha \text{,}
\label{ua}
\end{equation}%
\begin{equation}
\bar{e}^{\mu }=e^{\mu }\cosh \alpha -u^{\mu }\sinh \alpha  \label{ea}
\end{equation}%
where it follows from the conservation law that $\cosh \alpha =\frac{%
\varepsilon }{N}$ and it is chosen $\alpha >0$. Then, the transformation of
the curvature components under the local boosts to the orthonormal frame of
an infalling observer ($u^{1}<0$) read

\begin{equation}
\hat{R}_{1^{\prime }a^{\prime }1^{\prime }b^{\prime }}=\hat{R}_{1a1b}+\sinh
^{2}\alpha Z_{ab}\text{,}  \label{1a1b}
\end{equation}

\begin{equation}
\hat{R}_{a^{\prime }b^{\prime }c^{\prime }d^{\prime }}=\hat{R}_{abcd}\text{,}
\label{abcd}
\end{equation}%
\begin{equation}
\hat{R}_{1^{\prime }a^{\prime }b^{\prime }c^{\prime }}=\cosh \alpha \hat{R}%
_{1abc}\text{,}  \label{1abc}
\end{equation}%
\begin{equation}
\hat{R}_{0^{\prime }a^{\prime }b^{\prime }c^{\prime }}=-\sinh \alpha \hat{R}%
_{1abc}\text{,}  \label{0abc}
\end{equation}%
\begin{equation}
\hat{R}_{0^{\prime }a^{\prime }1^{\prime }b^{\prime }}=-\cosh \alpha \sinh
\alpha Z_{ab}\text{,}  \label{0a1b}
\end{equation}%
\begin{equation}
\hat{R}_{0^{\prime }1^{\prime }a^{\prime }b^{\prime }}=0\text{,}
\end{equation}%
\begin{equation}
\hat{R}_{0^{\prime }1^{\prime }1^{\prime }a^{\prime }}=\hat{R}_{010a}\sinh
\alpha \text{,}  \label{011}
\end{equation}

\begin{equation}
\hat{R}_{0^{\prime }1^{\prime }0^{\prime }1^{\prime }}=\hat{R}_{0101}\text{,}
\end{equation}%
\begin{equation}
\hat{R}_{0^{\prime }a^{\prime }0^{\prime }b^{\prime }}=\hat{R}_{0a0b}+\sinh
^{2}\alpha Z_{ab}
\end{equation}%
\begin{equation}
\hat{R}_{0^{\prime }1^{\prime }0^{\prime }a^{\prime }}=\hat{R}_{010a}\cosh
\alpha \text{,}  \label{010a}
\end{equation}

\begin{equation}
\bar{Z}_{ab}=Z_{ab}(2\cosh ^{2}\alpha -1)  \label{z0z}
\end{equation}%
where the combination 
\begin{equation}
Z_{ab}=R_{\mu a\nu b}(u^{\mu }u^{\nu }+e^{\mu }e^{\nu })  \label{zab}
\end{equation}%
and similarly for $\bar{Z}_{ab}$. One can check that (\ref{1a1b}) - (\ref%
{zab}) agree with formulas (\ref{tr1}), (\ref{tr2}), (\ref{z}) for the
spherically-symmetric case.

Using explicit formulas for the curvature tensor (see, e.g., the collection
of useful formulas in \cite{vis}, extended slightly in \cite{v}) we obtain
that%
\begin{equation}
\hat{R}_{1abc}=K_{ac;b}-K_{ab;c}\text{,}  \label{nabc}
\end{equation}%
\begin{equation}
\hat{R}_{010a}=\frac{R_{010a}}{N^{2}}=\frac{\partial
_{n}N_{;a}+K_{a}^{b}N_{;b}}{N}\text{,}  \label{0n0a}
\end{equation}%
\begin{equation}
Z_{ab}=\frac{N_{;a;b}}{N}-\frac{K_{ab}N^{\prime }}{N}+\frac{\partial K_{ab}}{%
\partial n}+(K^{2})_{ab}\text{.}  \label{z0}
\end{equation}%
Here (...)$_{;a}$ denotes covariant derivative with respect to the
two-dimensional metric $\gamma _{ab}$, the tensor of the extrinsic curvature 
$K_{ab}=-\frac{1}{2}\frac{\partial \gamma _{ab}}{\partial n}$, $%
(K^{2})_{ab}=K_{ac}K_{b}^{c}$. As an observer approaches the horizon, $\cosh
\alpha \rightarrow \infty $ but, typically, $Z_{ab}\rightarrow 0$ and the
behavior of the product in (\ref{z0z}) is not obvious in advance.

Let us denote $Z\equiv \frac{1}{2}Z_{ab}\gamma ^{ab}$. \ It follows from (%
\ref{z0}) that 
\begin{equation}
2Z=\frac{\Delta _{2}N}{N}+K^{\prime }-K\frac{N^{\prime }}{N}-SpK^{2}\text{, }%
K=K_{a}^{a}\text{, }SpK^{2}=K_{ab}K^{ab}\text{.}  \label{zk}
\end{equation}%
Further, one can check, using explicit formulas for the curvature tensor 
\cite{vis} that the combination (\ref{zk}) reduces to $G_{n}^{n}-G_{0}^{0}=8%
\pi (T_{n}^{n}-T_{0}^{0})\equiv 8\pi (\rho +p_{n})$ where $\rho =T_{\mu \nu
}u^{\mu }u^{\nu }$ is the energy density measured by a static observer, $%
p_{n}=T_{\mu \nu }e^{\mu }e^{\nu }$ is the corresponding pressure in the $n$%
-direction. Then, we obtain from (\ref{ua}), (\ref{ea}) that%
\begin{equation}
\bar{\rho}=T_{\mu \nu }\bar{u}^{\mu }\bar{u}^{\nu }=\frac{\bar{Z}+Z}{8\pi }%
-p_{n}\text{, }  \label{rzd}
\end{equation}%
which is the direct analog of eq. (\ref{rrz}) (with $L=0$ for radial motion)
which applies now to observers moving along geodesics in the $n$ direction
in the generic spacetime (\ref{gd}). The pressure transforms in a similar
way, so that%
\begin{equation}
\bar{p}_{n}=\frac{\bar{Z}+Z}{8\pi }-\rho \text{, }\bar{p}_{n}=T_{\mu \nu }%
\bar{e}^{\mu }\bar{e}^{\nu }
\end{equation}%
that is equivalent to $\tilde{Z}=Z\,(2\frac{\varepsilon ^{2}}{U}-1)$ in
accordance with (\ref{z}).

It is clearly seen from (\ref{rzd}) that in the generic case of distorted
horizons $\bar{\rho}$ (energy density) and $\bar{Z}$ (tidal forces) are
either both infinite or both finite in a free-falling frame. (The same
statement concerns also the relationship between the pressure $\bar{p}_{n}$
and $\bar{Z})$. This applies to non-extremal, extremal or ultraextremal
horizons and enlarges on the observation made in \cite{v} due to explicit
examining different asymptotical behavior of the metric near the horizon. In
the spherically-symmetric case we return to (\ref{rrz}). Apart from this,
one should take into account the behavior of components (\ref{1abc}), (\ref%
{0abc}), (\ref{011}), (\ref{010a}) which were absent for
spherically-symmetrical metrics.

Below we discuss different types of horizons separately.

\subsection{Non-extremal case}

It follows from the finiteness of the Kretschmann scalar on the horizon
requires that the relevant metric functions have the asymptotic expansions 
\cite{vis}

\begin{equation}
N=\kappa _{H}n+\frac{\kappa _{2}(x^{a})}{3!}n^{3}+\frac{\kappa _{3}(x^{a})}{%
4!}n^{4}+O(n^{5})\text{,}
\end{equation}%
\begin{equation}
\gamma _{ab}=[\gamma _{H}]_{ab}(x^{a})+\frac{[\gamma _{2}]_{ab}n^{2}}{2!}+%
\frac{[\gamma _{3}]_{ab}n^{3}}{3!}+O(n^{4})  \label{gn}
\end{equation}%
\begin{equation}
K_{ab}=K_{ab}^{(1)}n+\frac{K_{ab}^{(2)}}{2!}n^{2}+\frac{K_{ab}^{(3)}}{3!}%
n^{2}+O(n^{4})\text{, }K_{ab}^{(1)}=-\frac{[\gamma _{2}]_{ab}}{2}\text{, }%
K_{ab}^{(2)}=-\frac{[\gamma _{3}]_{ab}}{2}\text{,}  \label{k}
\end{equation}%
$n$ is the proper distance from the horizon, the constant $\kappa _{H}$ has
the meaning of the surface gravity. Then, one obtains that near the horizon%
\begin{equation}
Z_{ab}=\frac{K_{ab}^{(2)}}{2}n+O(n^{2})\text{,}
\end{equation}%
so%
\begin{equation}
\bar{Z}_{ab}=\frac{\varepsilon ^{2}}{\kappa _{H}^{2}}%
K_{ab}^{(2)}n^{-1}+Y_{ab}
\end{equation}%
where $Y_{ab}=const$. It follows from (\ref{1abc}), (\ref{0n0a}) that $\hat{R%
}_{1abc}\sim \hat{R}_{010a}\sim n.$ As $\cosh \alpha \sim \sinh \alpha \sim
N^{-1}\sim n^{-1}$, we obtain that in the boosted frame the components $\ 
\hat{R}_{1^{\prime }a^{\prime }b^{\prime }c^{\prime }}$, $\hat{R}_{0^{\prime
}a^{\prime }b^{\prime }c^{\prime }}$, $\hat{R}_{0^{\prime }1^{\prime
}1^{\prime }a^{\prime }}$ and $\hat{R}_{0^{\prime }1^{\prime }0^{\prime
}a^{\prime }}$ are finite and in general non-vanishing. The component $\hat{R%
}_{0^{\prime }a^{\prime }1^{\prime }b^{\prime }}$ behaves like $\bar{Z}_{ab}$
(the latter applies to other types of the horizons as it is seen from (\ref%
{0a1b}) and (\ref{z0z})).

In the spherically-symmetric case $r-r_{0}\sim n^{2}$, so that the expansion
(\ref{k}) contains only odd powers, $K_{ab}^{(2)}=0$ and $\bar{Z}_{ab}$ is
finite. Thus, spherically-symmetric non-extremal TNBH do not exist in
accordance with table 1, line "$p=q=1$"$.$ However, simply naked black holes
are possible in accordance with \cite{nk1}, \cite{nk2}.

\subsection{Ultraextremal case}

Now we consider the metric which reads \cite{v}

\begin{equation}
N=\frac{A_{1}(x^{a})}{n^{m}}+\frac{A_{2}(x^{a})}{n^{m+1}}+O(n^{-m-2})\text{, 
}
\end{equation}%
\begin{equation}
\gamma _{ab}=\gamma _{ab}^{(0)}+\frac{\gamma _{^{(1)}ab}}{n^{s}}+O(n^{-s-1})%
\text{, }s>0\text{, }m>0\text{,}
\end{equation}%
\begin{equation}
K_{ab}=\frac{s\gamma _{^{(1)}ab}}{2n^{s+1}}+O(n^{-s-2})\text{,}
\end{equation}%
the horizon is at infinite proper distance, the Kretschmann scalar is finite
on the horizon. In the spherically-symmetrical case it reduces to (\ref{pq})
with%
\begin{equation}
p=2+\frac{2}{s}>2\text{, }q=\frac{m}{s}\text{.}
\end{equation}%
\begin{equation}
s=\frac{2}{p-2}\text{, }m=\frac{2q}{p-2}\text{, }p>2\text{.}
\end{equation}%
Then, previous consideration applies - see table 1.

In the distorted case one finds that on the horizon%
\begin{equation}
Z_{ab}=\frac{A_{1a;b}}{A_{1}}
\end{equation}%
Then, if $\left( A_{1}\right) _{;a}\neq 0$, 
\begin{equation}
\bar{Z}_{ab}\sim \frac{Z_{ab}}{N^{2}}\sim n^{2m}\rightarrow \infty \text{.}
\end{equation}

The crucial point is that this contribution comes from the terms which were
absent in the spherically-symmetrical case but dominate now. In doing so, $%
\hat{R}_{1abc}\sim n^{-s-1}$, $\hat{R}_{010a}\sim n^{-1}$, $\hat{R}%
_{0^{\prime }1^{\prime }0^{\prime }a^{\prime }}\sim n^{m-1}\sim \hat{R}%
_{0^{\prime }1^{\prime }1^{\prime }a^{\prime }}$, $\hat{R}_{1^{\prime
}a^{\prime }b^{\prime }c^{\prime }}$, $\hat{R}_{0^{\prime }a^{\prime
}b^{\prime }c^{\prime }}$ $\sim n^{m-s-1}$. All such black holes are truly
naked in agreement with \cite{v}. But even if $A_{1}=const$, this does not
guarantee the absence of TNBH. Indeed, in this case we must take into
account first corrections. Let $A_{1}$ is a constant but $A_{2}$ is not. The
first term in r.h.s. of (\ref{z0}) dominates and, as a result, 
\begin{equation}
Z_{ab}\sim n^{-1}\text{.}
\end{equation}%
Then, 
\begin{equation}
\bar{Z}_{ab}\sim \frac{Z_{ab}}{N^{2}}\sim n^{2m-1}\text{.}
\end{equation}%
If $m>\frac{1}{2}$ tidal forces diverge in the boosted frame.

Let us consider now the behavior of the rest of components. We have $\cosh
\alpha \sim \sinh \alpha \sim n^{m}$, $\hat{R}_{1abc}\sim n^{-s-1}$, $\hat{R}%
_{010a}\sim n^{-2}$. As a result, $\hat{R}_{0^{\prime }1^{\prime }0^{\prime
}a^{\prime }}\sim \hat{R}_{0^{\prime }1^{\prime }1^{\prime }a^{\prime }}\sim
n^{m-2}$, $\hat{R}_{1^{\prime }a^{\prime }b^{\prime }c^{\prime }}$, $\hat{R}%
_{0^{\prime }a^{\prime }b^{\prime }c^{\prime }}$ $\sim n^{m-s-1}$ and
diverge if $m>s+1$. This criterion is more tight than for tidal transverse
forces. Thefore, it is the behavior of $\bar{Z}_{ab}$ that forces us to
include a blackhole into the class of TNBHs. We have usual ($m<\frac{1}{2}$%
), naked ($m=\frac{1}{2}$) or truly naked ($m>\frac{1}{2}$) black hole.

\subsection{Extremal case, $p=2$}

Consider another typical case that corresponds to the finite Kretschmann
scalar on the horizon \cite{v}:

\begin{equation}
N=B_{1}(x^{a})\exp (-\frac{n}{n_{0}})+B_{2}(x^{a})\exp (-\frac{2n}{n_{0}}%
)+B_{3}(x^{a})\exp (-\frac{3n}{n_{0}})+...\text{, }
\end{equation}%
$n_{0}>0$ is a constant,%
\begin{equation}
\gamma _{ab}=\gamma _{ab}^{(0)}+\gamma _{^{(1)}ab}\exp (-\frac{n}{n_{0}}%
)+O(\exp (-\frac{2n}{n_{0}}))\text{,}  \label{ge}
\end{equation}%
$n\rightarrow \infty $.

Then, near the horizon%
\begin{equation}
Z_{ab}=\frac{1}{2}\lim_{n\rightarrow \infty }\frac{N_{;a;b}}{N}\text{,. }%
\bar{Z}_{ab}\sim \frac{Z_{ab}}{N^{2}}\text{.}
\end{equation}%
If a) $(B_{1})_{;a}\neq 0$, it turns out that $Z_{ab}\neq 0$ is finite and $%
\bar{Z}_{ab}$ is infinite (TNBH) and diverges like $N^{-2}$. If b) $%
(B_{1})_{;a}=0$, $(B_{2})_{;a}\neq 0$, $Z_{ab}=0$ but $\bar{Z}_{ab}$ is
still infinite (TNBH) and diverges like $N^{-1}$. If c) $%
(B_{1})_{;a}=0=(B_{2})_{;a}$, $(B_{3})_{;a}\neq 0$, $Z_{ab}=0$, $\bar{Z}%
_{ab} $ is finite (naked black hole). If d) $%
(B_{1})_{;a}=(B_{2})_{;a}=(B_{3})_{;a}=0$, $(B_{4})_{;a}\neq 0$, $Z_{ab}=0$ $%
=\bar{Z}_{ab}$ (usual).

As far as the other curvature components is concerned, $\hat{R}_{1abc}\sim
\exp (-\frac{n}{n_{0}})\sim N$, so that $\hat{R}_{1^{\prime }a^{\prime
}b^{\prime }c^{\prime }}$, $\hat{R}_{0^{\prime }a^{\prime }b^{\prime
}c^{\prime }}$ are finite due to the compensating factor $\cosh \alpha $ or $%
\sinh \alpha $ $\sim N^{-1}$ in (\ref{1abc}), (\ref{0abc}). However, the
component $\hat{R}_{010a}$ is now non-vanishing in general, according to (%
\ref{0n0a}), so it follows from (\ref{011}) and (\ref{010a}) that in the
boosted frame the components $\hat{R}_{0^{\prime }1^{\prime }0^{\prime
}a^{\prime }}$ and $\hat{R}_{0^{\prime }1^{\prime }1^{\prime }a^{\prime }}$
diverge like $N^{-1}$. It is weaker than for $\bar{Z}_{ab}$ in case a) and
is the same in case b). We see again, that the behavior of tidal forces in
the transverse direction is sufficient to conclude that the object belongs
to the class of TNBHs.

\section{Summary and discussion}

In practice, the regularity of the energy encountered by a free-falling
observer, is often considered as a criterion of the regularity of spacetime
at the horizon. In particular, the typical test exploited for examining the
existence (or nonexistence) of quantum-corrected extremal or ultraextremal
black holes consists in determining whether or not this quantity is finite 
\cite{qb}. Meanwhile, we saw that, actually, there are three different
criteria: (i) the finiteness of the Kretschmann scalar which is a standard
condition of the regularity of the geometry, (ii) the finiteness of the
energy density, (iii) the finiteness of separate curvature components in a \
given frame (in particular, the finiteness of tidal forces). We saw that the
conditions (ii) and (iii) are equivalent to each other but, in general, they
are not equivalent to (i). More precisely, if we are interested in the
spacetimes regular in the sense (i) only, it entails the validity of (ii)
and (iii) for a static observer. However, for a free-falling observer (ii)
and (iii) may be violated without violation of (i).

The corresponding object called "truly naked black holes" (TNBH) is the
ultimate extension of "naked black holes" of Refs.\cite{nk1}, \cite{nk2} for
which tidal forces experienced by a free-falling observer are enhanced with
respect to the static frame but remain finite. This extension became
possible due to consideration or spherically-symmetrical ultra-extremal and
extremal black holes or distorted horizons. The class of these objects
includes also examples with the infinite area of a horizon found earlier for
particular cases in \cite{br1}, \cite{br2}. In the present article we also
showed that TNBH may be spherically-symmetrical, not only distorted \cite{v}%
. As far as the structure of the Riemann tensor is concerned, now
divergencies in the free-falling frame are not due to Weyl scalars (as it
was for distorted TNBH considered in \cite{v}) but entirely due to the Ricci
tensor (or corresponding energy density $\bar{\rho}$ and pressure $\bar{p}%
_{n}$). In principle, in the most general case divergencies may occur in
both types of quantities. We also found which components are enhanced to
infinity and which remain finite. In particular, for distorted TNBH some new
diverging components appear that have no analogue in the
spherically-symmetrical case.

From a more general viewpoint, the objects discussed in our paper are
intimately connected with so-called non-scalar polynomial curvature
singularities \cite{he} (Chap. 8), \cite{es} (Sec. 4.1, 4.2). They appeared
when there are no diverging scalar fields but components of the curvature
tensor in some frames behave badly. This happens if local Lorentz boosts
from one frame to another also behave badly. In the present context, a
static observer which resides near the horizon, becomes badly determined in
the near-horizon limit. The force which is needed to support him in the
rest, grows unbound and, roughly speaking, a time-like observer tends to a
light-like one. Correspondingly, the local Lorentz boost from a static frame
to a free-falling one, becomes ill-defined, its angle parameter $\cosh
\alpha $ diverges in the horizon limit. Thus, black hole physics supplies us
with the mechanism in which the singularities under discussion are generated
in a natural way. However, by itself the presence of \ a horizon does not
lead to such singularities. It depends also on the rate under which the
boost parameter diverge and details of the asymptotic behavior of the
curvature tensor near the horizon. Depending on these properties, one
obtains usual black holes (like the Schwarzschild or the Reissner-Norstr\"{o}%
m ones) where there is no enhancement of the curvature components at all,
their enhancement from zero to finite values (like in naked black holes of 
\cite{nk1}, \cite{nk2}) or truly naked black holes considered in our paper.
Only in the third case one can speak about singularities in the above sense.

The existence of TNBH configurations discussed in the present article as
well as in the previous one \cite{v} points to some potential rooms in
scenarios of gravitational collapse which need further consideration. It
also hints that the cosmic censorship should be somehow reformulated to take
into account these subtleties. It was pointed out in \cite{nk1} that the
existence of naked black hole may affect the issue of information loss and
black hole entropy since large tidal forces disturb significantly the matter
falling into a black hole. The more so, this factor becomes important in the
case of TNBH when tidal forces are not simply large but infinite on the
horizon.

We examined the non-extremal, extremal and ultra-extremal types of a horizon
and found that the situation when tidal forces in the transverse direction
are finite but other curvature components diverge is impossible. Therefore,
actually it is behavior of tidal transverse forces that enables us to
classify an object as TNBH. We also demonstrated that in the
spherically-symmetrical case such a black hole should be extremal or
ultra-extremal. In the latter case the null energy condition may be violated
in some vicinity of a horizon in the outward direction.

To summarize, there are three types of horizons in the aspect under
discussion: "usual" (in both frames curvature components are finite, tidal
forces are zero), naked (in both frames curvature components are finite,
tidal forces are zero for a static observer but finite non-vanishing for a
free-falling observer), "truly naked" (some curvature components are
infinite for a free-falling observer). In the context of the backreaction
problem, all examples analyzed in \cite{qb} fall into the first class ($p=q=2
$ for the extremal subcase and $p=q=3$ for the ultraextremal one) in the
unperturbed case. However, according to line 6 of Table I, corresponding
quantum-corrected metrics represent naked black holes, so in this sense
quantum backreaction is able to change the type of an extremal or
ultraextremal horizon.

\begin{acknowledgments}
I thank Vojtech Pravda for helpful comments.
\end{acknowledgments}


\begin{thebibliography}{99}
\bibitem{nk1} G. T. Horowitz and S. F. Ross, Phys. Rev. D \textbf{56}, 2180
(1997).

\bibitem{nk2} G. T. Horowitz and S. F. Ross, Phys. Rev. D \textbf{57}, 1098
(1998).

\bibitem{br1} K.A. Bronnikov, G. Cl\'{e}ment, C.P. Constantinidis, J.C.
Fabris, Grav.Cosmol. \textbf{4 }128 (1998).

\bibitem{br2} K.A. Bronnikov, G. Cl\'{e}ment, C.P. Constantinidis, J.C.
Fabris, Phys.Lett. A \textbf{243 }121 (1998).

\bibitem{v} V. Pravda and O. B. Zaslavskii, Class. Quant. Grav. \textbf{22},
5053 (2005).

\bibitem{he} S. W.\ Hawking and G.\ F. Ellis, \textit{Large scale structure
of Universe }(Cambridge, Cambridge University Press, 1973).

\bibitem{es} G. F. R. Ellis and B. G. Schmidt, Gen. Rel. Grav. \textbf{8},
913 (1977).

\bibitem{w} R M Wald, \textit{General Relativity} (Chicago, IL: Chicago
University Press, 1984).

\bibitem{2d} D. J. Loranz, W. A. Hiscock and P. R. Anderson, Phys. Rev. D 
\textbf{52}, 4554 (1995).

\bibitem{gap} P. R. Anderson and C. D. Mull, Phys. Rev. D \textbf{59},
044007 (1999).

\bibitem{rom} L. J. Romans, Nucl. Phys. \textbf{B} 383, 395 (1992).

\bibitem{vis} A. J. M. Medved, D. Martin and M. Visser Class, Quantum Grav. 
\textbf{21} 3111 (2004).

\bibitem{ac1} A. Riess et al., Astron. J. 116, 1009 (1998).

\bibitem{ac2} S. J. Perlmutter et al., Astrophys. J. 517, 565 (1999).

\bibitem{th} M. S. Moris and K. S. Thorne, Am. J. Phys. \textbf{56,} 395
1988).

\bibitem{book} M. Visser, \textit{Lorentzian Wormholes: From Einstein to
Hawking }(AIP Press,New York, 1995).

\bibitem{ph} O. B. Zaslavskii, Phys. Rev. \textbf{D} 72, 061303 (2005).

\bibitem{j} Jordan P Z. Phys. Rev. \textbf{157,} 112 (1959).

\bibitem{bd} Brans C H and Dicke R H Phys. Rev. \textbf{124,} 925 (1961)..

\bibitem{qb} P. R. Anderson, W. A. Hiscock and D. A. Samuel, Phys. Rev.
Lett. \textbf{70} (1993) 1739; J. Matyjasek and O. B. Zaslavskii, Phys. Rev.
D \textbf{64}, 104018 (2001); Jerzy Matyjasek, O. B. Zaslavskii, Phys.Rev. D 
\textbf{71 }(2005) 087501; Arkady A. Popov and O. B. Zaslavskii, Phys. Rev.
D \textbf{75}, 084018 (2007).

\bibitem{cl} M. Campanelli and C.O. Lousto, Int. J. Mod. Phys. \textbf{D} 2,
451 (1993)
\end{thebibliography}
\end{document}